Article

# Dynamic Capabilities for AI-Enabled Exploration: Antecedents, Mechanisms, and Innovation Outcomes

**Thabit Atobishi [1,*] and Saeed Nosratabadi [2]**

[1] Department of Health Information Management and Technology, College of Applied Medical Sciences, King Faisal University, Al-Ahsa 31982, Saudi Arabia
[2] Rawls College of Business, Texas Tech University, Lubbock, TX 79409, USA; snosrata@ttu.edu
* Correspondence: tatubaishi@kfu.edu.sa

**Abstract**

While the operational benefits of Artificial Intelligence (AI) are well-documented, the mechanisms through which firms leverage AI for strategic exploration and radical innovation remain under-theorized. This study addresses the "black box" of AI value creation by integrating the Technology–Organization–Environment (TOE) framework with the Dynamic Capabilities View (DCV). We propose that AI adoption is not a direct antecedent to performance but a multi-stage process wherein technological, organizational, and environmental factors enable the development of sensing capability, which in turn fosters a novel capability we term "AI-Enabled Exploration." Analyzing survey data from 245 senior executives in Saudi Arabia, a high-growth economy undergoing state-led digital transformation, we employed Partial Least Squares Structural Equation Modeling (PLS-SEM) to test the model. The results confirm a serial mediation chain: organizational readiness and technology compatibility drive sensing capability, which subsequently powers AI-enabled exploration to enhance innovation performance. Contrary to expectations, government support was not a significant predictor of sensing capability, suggesting that in resource-rich environments, external incentives are necessary but insufficient for capability building. Furthermore, competitive pressure was found to positively moderate the relationship between organizational readiness and exploration, acting as a critical catalyst that converts latent resources into active experimentation. These findings offer a theoretical roadmap for firms attempting to transition from AI-driven efficiency to AI-driven ambidexterity.

## 1. Introduction

The landscape of modern business operations has been profoundly transformed by the continuous advancement of digital technologies, with artificial intelligence (AI) emerging as a foundational element for competitive survival [1,2]. While early iterations of AI were primarily deployed to automate routine tasks and enhance operational efficiency and AI adoption can positively impact the decision-making process by enabling more accurate and timely decisions [3], which is a well-documented phenomenon in recent literature [4,5], the advent of Generative AI has introduced a new strategic imperative. Firms are no longer just looking to AI for exploitation, the refinement of existing competencies and cost reduction [6], but are increasingly exploring its potential for exploration, or the discovery of novel business models, products, and market opportunities [7,8].

This shift raises an important question about the nature of AI-driven exploration: does the initiative come from organizational leaders who use AI as a tool to develop and test their own strategic ideas, or does AI increasingly function as a source of novel directions that leaders then evaluate and act upon? Kollmann et al. frame this as the central challenge of Artificial Leadership, positioning digital transformation as a leadership task at the boundary between human agency and artificial intelligence [9]. Our study addresses this question empirically by examining the organizational capability conditions under which AI transitions from supporting human-directed processes to generating new strategic possibilities for leaders to mobilize around.

Despite the transformative potential of AI, a significant theoretical gap remains regarding how organizations can systematically leverage these technologies for exploration. The majority of Information Systems (IS) research has focused on the "efficiency" dividend of AI, examining how analytics capabilities drive operational performance and supply chain agility [10–12]. For instance, Wong and Ngai (2025) recently demonstrated how dynamic capabilities mediate the relationship between technological factors and operational performance in the telecommunications sector [5]. However, the mechanisms through which AI enables firms to break away from path-dependent routines and engage in high-risk, high-reward exploration remain under-theorized [13,14]. This "black box" is critical, as the capabilities required to *sense* new opportunities via AI are distinct from those used to optimize existing processes [15,16].

Furthermore, the environmental context of AI adoption plays a pivotal role that is often oversimplified in single-country studies [16, 17]. Most empirical evidence stems from Western contexts where innovation is market-driven. In contrast, emerging high-growth economies like the Kingdom of Saudi Arabia (KSA) present a unique "leapfrog" environment. Under the Saudi Vision 2030 framework, organizations face immense state-led pressure to digitize, coupled with abundant financial resources but varying levels of organizational readiness [18,19]. This creates a distinct tension: firms possess the resources to adopt sophisticated AI, but may lack the dynamic capabilities to translate these assets into genuine exploratory innovation rather than mere automation [20].

To address these gaps, this study integrates the Technology–Organization–Environment (TOE) framework [21] with the Dynamic Capabilities View (DCV) [22]. We posit that AI adoption is not a monolithic event but a process that feeds specific dynamic capabilities. We introduce and provide preliminary empirical evidence for a novel construct, "AI-Enabled Exploration," defined as the organizational capability to utilize AI systems for experimenting with new business approaches and inventing novel products [23]. By linking TOE antecedents to this specific capability, and subsequently to innovation performance, we offer a more granular understanding of AI's strategic value.

Consequently, this study addresses the following research questions:

1. What technological, organizational, and environmental (TOE) factors influence the development of a firm's AI-related dynamic capabilities (i.e., Sensing Capability and AI-Enabled Exploration)?
2. What is the mediating role of these dynamic capabilities in the relationship between TOE factors and a firm's innovation performance?
3. How does competitive pressure, as an environmental factor, moderate the relationship between organizational resources and the development of AI-Enabled Exploration?

This study makes three primary contributions to the IS literature. First, it extends the work of Wong and Ngai (2025) by shifting the dependent variable from operational performance to innovation performance [5], thereby addressing the "exploration" side of the ambidexterity dilemma [24]. Second, it conceptualizes AI-Enabled Exploration as a distinct dynamic capability, providing empirical evidence that AI is an antecedent to strategic renewal, not just efficiency. Third, it provides rare empirical insights from the Middle East, enriching the global discourse on digital transformation in resource-rich, state-guided economies.

## 2. Theoretical Background

### *2.1. The Dynamic Capabilities View (DCV)*

The Dynamic Capabilities View (DCV) serves as the overarching theoretical lens for this study. Teece et al. defined dynamic capabilities as the firm's ability to integrate, build, and reconfigure internal and external competencies to address rapidly changing environments [25]. Unlike ordinary capabilities (which focus on operational efficiency or "doing things right"), dynamic capabilities focus on strategic renewal (or "doing the right things"). Dynamic capabilities play an important role in enhancing the competitive position and improving the overall performance of the organizations [26].

Within the IS literature, dynamic capabilities are often categorized into specific dimensions: sensing (identifying opportunities), seizing (mobilizing resources), and transforming (reconfiguring assets) [15]. Kowalski et al. further unpack the microfoundations of these capabilities in the context of digital transformation, highlighting the role of data-driven culture and digital analytics in enabling firms to sense and respond to market changes [27]. AI is not merely a technical asset but an enabler of these dynamic capabilities [11]. This study addresses all three dimensions, though not with equal theoretical depth. Sensing capability is the primary focus because it represents the understudied foundational mechanism through which AI adoption translates into strategic action. AI-Enabled Exploration, however, is conceptualized as a seizing capability: it captures how firms mobilize their AI assets to actively pursue the opportunities they have identified. Innovation performance, in turn, reflects the transforming outcome, capturing how firms reconfigure their knowledge and assets into new products and market responses. This sequential mapping of sensing, seizing, and transforming onto the model's three core constructs is consistent with Teece's micro-foundational logic [22]. The emphasis on sensing is theoretically justified because in resource-rich, state-guided economies such as Saudi Arabia, the primary bottleneck in AI value creation lies not in resource availability but in the organizational capacity to identify which opportunities are worth pursuing.

### *2.2. AI-Enabled Exploration and Ambidexterity*

Organizational learning theory distinguishes between exploitation, refining existing knowledge, and exploration, searching for new knowledge beyond the firm's current domain [8]. While early AI adoption focused primarily on exploitation through automation and cost reduction, the emergence of Generative AI has shifted this trajectory toward exploration. This shift introduces a new form of organizational capability that warrants precise theoretical conceptualization. We define "AI-Enabled Exploration" as a specific dynamic capability wherein the organization utilizes AI algorithms to simulate novel scenarios, generate creative content, and identify adjacent market opportunities that lie outside the firm's current experience scope [23]. This capability represents the mechanism through which digital assets are converted into strategic renewal.

AI-Enabled Exploration shares conceptual proximity with several established constructs in the IS and management literature, specifically exploratory innovation, organizational ambidexterity, and digital innovation capability. These constructs appear similar on the surface because they all concern organizational efforts to pursue novelty and renewal. However, each differs from AI-Enabled Exploration in meaningful ways, and conflating them would obscure the distinct theoretical contribution of this study.

Exploratory innovation, as defined by Jansen et al., refers to the organizational pursuit of new knowledge, products, and markets that lie outside the firm's current experience domain [23]. It captures the behavioral tendency of firms to experiment with novel ideas and depart from established routines. As an outcome-oriented construct, exploratory innovation describes what a firm does rather than how it does it. This is precisely where it diverges from AI-Enabled Exploration. A firm can engage in exploratory innovation through traditional R&D or human creativity alone. AI-Enabled Exploration, by contrast, is specifically enabled by generative algorithms and simulation tools that allow firms to prototype and experiment digitally at scale. AI-Enabled Exploration is therefore best understood as an antecedent to exploratory innovation, the capability that makes large-scale exploration feasible, rather than a synonym for it.

Organizational ambidexterity, defined by Raisch and Birkinshaw, refers to the firm's capacity to simultaneously pursue exploration and exploitation, balancing short-term efficiency with long-term renewal [7]. Operating at the strategic and structural level, it describes how organizations manage the inherent tension between refining existing competencies and searching for new ones. He and Wong empirically demonstrated that firms capable of balancing both modes consistently outperform those focused exclusively on either [24], and Hwang et al. further confirmed that ambidexterity mediates the relationship between innovation strategies and firm performance [28]. The key distinction from AI-Enabled Exploration lies in scope. Ambidexterity is a higher-order organizational design concept that encompasses both exploration and exploitation at once. AI-Enabled Exploration addresses only the exploration side of this balance, serving as a micro-level capability that contributes to the exploration component of ambidexterity without addressing exploitation. In this sense, AI-Enabled Exploration is a building block of ambidexterity rather than a substitute for it.

Digital innovation capability, as conceptualized by Nambisan et al., refers to the broad organizational capacity to harness digital technologies for the creation of new products, services, and business models [29]. This construct encompasses a wide range of digital tools and processes, from cloud computing to data analytics, and applies across all stages of the innovation process. Its generality is both its strength and its limitation as an explanatory construct. AI-Enabled Exploration differs in two important ways. First, it is narrower in scope, focusing specifically on generative AI and simulation tools rather than the full spectrum of digital technologies. Second, it is more precisely bounded in application, targeting the experimentation and prototyping phase rather than the entire innovation lifecycle. This specificity makes AI-Enabled Exploration more precisely measurable and actionable for organizations seeking to understand how AI specifically, rather than digital technology broadly, contributes to strategic renewal.

The role of AI-Enabled Exploration can also be situated within the Artificial Leadership framework proposed by Kollmann et al., which positions digital transformation as a leadership task at the intersection of human and artificial decision-making [9]. Kollmann et al. distinguish between AI that operates 'for' human leaders, supporting data-driven operational processes, and AI that increasingly operates 'to' human leaders, generating strategic directions that decision-makers then evaluate and execute. This distinction maps directly onto the two-stage mechanism we propose. Sensing Capability reflects the 'AI for human' mode: leaders deploy AI tools to scan the environment, interpret market signals, and identify opportunities. AI-Enabled Exploration captures the transition toward the 'AI to human' mode: generative algorithms and simulation tools produce novel scenarios and prototypes that leaders then mobilize organizational resources around. Kollmann et al. further situate this challenge within a framework of digital ambidexterity, distinguishing between exploitation-oriented AI deployment, which supports existing processes, and exploration-oriented AI deployment, which generates new possibilities. Our study specifically addresses the exploration side of this framework, contributing empirical evidence about the organizational conditions that determine whether AI serves firms as an efficiency tool or as a genuine source of exploratory innovation.

### 2.3. The TOE Framework

To delineate the antecedents of AI-enabled capabilities, we adopt the Technology–Organization–Environment (TOE) framework proposed by Tornatzky and Fleischer [21]. The TOE framework has garnered widespread empirical support within the IS literature for its robust explanatory power in analyzing firm-level technology adoption and assimilation, including its recent application to digital transformation and AI adoption decisions in small and micro enterprises [30,31]. Unlike individual-level acceptance models, the TOE framework emphasizes that technological outcomes are not determined solely by the technology itself but are shaped by a confluence of internal firm characteristics and external environmental forces. In this study, we utilize the framework to categorize the critical drivers that enable a firm to develop sophisticated sensing mechanisms.

The framework identifies three distinct contexts that collectively influence organizational outcomes. The technological context describes the internal and external technologies relevant to the firm, focusing in this study on Technology Compatibility, which assesses the degree to which AI systems align with existing technical infrastructure and values to minimize implementation friction. The organizational context refers to the characteristics and resources of the firm, captured here by Organizational Readiness; this encompasses the availability of financial capital and specialized technical skills required to deploy and manage complex AI tools. The environmental context refers to the arena in which the firm conducts its business, including the regulatory landscape and industry structure. Specifically, we examine Government Support and Competitive Pressure as the primary external forces that incentivize or constrain AI capability development.

Prior IS research has consistently shown that these TOE contexts collectively enable firms to build dynamic capabilities. Wong and Ngai demonstrated that technological and organizational factors serve as antecedents to dynamic capabilities rather than direct drivers of performance [5]. Similarly, Mikalef and Gupta showed that organizational factors shape AI capability development, which subsequently drives firm performance [11]. Zhu et al. further confirmed that TOE factors determine the depth of technology assimilation within firms [32]. Collectively, this body of evidence positions TOE factors as the foundational conditions through which dynamic capabilities are cultivated.

Building on this established foundation, we argue that TOE factors specifically enable the development of Sensing Capability, which refers to the firm's ability to identify market opportunities. We posit sensing capability as the necessary condition that activates AI-Enabled Exploration: without first identifying opportunities, AI systems cannot be effectively deployed for experimentation. It is this sensing-to-exploration mechanism, not the TOE-DCV integration itself, that constitutes the core theoretical contribution of this study.

### 2.4. Research Model and Hypotheses

Drawing on the theoretical integration of the TOE framework and the Dynamic Capabilities View (DCV), we posit that technological, organizational, and environmental factors serve as the foundational antecedents that enable a firm to build Sensing Capability [11]. We argue that this sensing mechanism acts as a critical precursor to AI-Enabled Exploration [22], which subsequently drives Innovation Performance. Furthermore, incorporating contingency theory [31], we propose that the translation of organizational resources into exploration behavior is contingent upon the level of Competitive Pressure in the external environment.

### 2.5. Determinants of Sensing Capability

The first part of our model examines the antecedents of sensing capability. According to the DCV, sensing is not an inherent trait but an organizational skill that must be built using specific resources and favorable environmental conditions [15].

We first consider Government Support, a critical environmental factor in state-guided economies. In the context of Saudi Arabia's Vision 2030, the government actively reduces the barriers to entry for advanced technologies through incentives and regulatory frameworks. Institutional theory suggests that such external support legitimizes new technologies and provides the necessary slack resources for firms to experiment with data gathering [33,34]. When firms perceive strong government backing, they are more likely to invest in the expensive, high-risk AI infrastructure required to scan global markets. Previous studies have consistently found that government mandates and incentives significantly reduce the perceived risks of adopting complex IS innovations, thereby facilitating capability development [35,36].

**H1.** *Government Support is positively associated with Sensing Capability.*

From the organizational context, Organizational Readiness represents the firm's reservoir of financial and technical resources. Drawing on the Resource-Based View (RBV), complex capabilities like sensing require the integration of tangible assets (infrastructure) and intangible assets (technical skills) [37]. AI-driven sensing is computationally intensive and requires specialized talent to interpret unstructured data streams. Iacovou et al. argued that without sufficient financial and technological readiness, firms lack the absorptive capacity to implement advanced inter-organizational systems [38], where absorptive capacity encompasses the firm's dynamic ability to acquire, assimilate, and transform external knowledge into innovation outcomes [39]. Recent AI literature supports this, noting that firms with high readiness possess the competence to deploy analytical tools effectively, allowing them to develop a deep, real-time awareness of the business environment [40].

**H2.** *Organizational Readiness is positively associated with Sensing Capability.*

Within the technological context, Technology Compatibility refers to the degree to which AI systems align with the firm's existing IT infrastructure and business values. Diffusion of Innovation (DOI) theory posits that incompatibility creates friction; if AI systems do not integrate with legacy databases, the firm spends its energy on technical fixes rather than market analysis [41]. Conversely, high compatibility ensures a seamless flow of data across the organization. This interoperability is essential for sensing capability, as it allows AI algorithms to aggregate internal and external data without bottlenecks, thereby providing a clearer and faster picture of market opportunities [36,42].

**H3.** *Technology Compatibility is positively associated with Sensing Capability.*

### 2.6. The Link Between Sensing and AI-Enabled Exploration

The core of our theoretical contribution lies in the relationship between sensing and exploration. While sensing capability allows a firm to identify opportunities, exploration capability involves the active experimentation and search for solutions to capture those opportunities [8]. We argue that sensing is a necessary condition for effective exploration.

According to Teece, dynamic capabilities are micro-founded on the sequence of sensing, seizing, and reconfiguring [22]. This sequential interplay has been empirically validated in emerging technology contexts, where sensing capabilities positively influence seizing and transforming capabilities, which in turn drive innovation performance [43]. AI-Enabled Exploration involves using generative models and simulations to prototype novel products and enter adjacent markets. However, exploration that is disconnected from market reality often leads to failure. Sensing capability provides the high-fidelity data "inputs" required for AI models to generate valuable "outputs" [44]. For instance, a firm that effectively senses a shift in consumer sentiment can use Generative AI to rapidly prototype products that match that specific shift. Therefore, consistent with the DCV logic that information acquisition precedes strategic action, we posit that firms with superior sensing capabilities are better positioned to deploy AI for exploration [45].

**H4.** *Sensing Capability is positively associated with AI-Enabled Exploration.*

This sensing-to-exploration sequence represents the deliberate, baseline mechanism through which organizational resources translate into exploratory behavior under normal environmental conditions, though as theorized in H6, this sequential process may be bypassed when competitive pressure is high.

### *2.7. Exploration and Innovation Performance*

The ultimate outcome variable in our model is Innovation Performance, defined as the success of new product introductions and the speed of market response. However, AI innovation tends to be less radical and more process-oriented compared to other IT innovations [46]. Organizational learning literature suggests that while exploitation leads to short-term efficiency, it is exploration that drives long-term radical innovation [13,24].

AI-Enabled Exploration allows firms to decouple the cost of experimentation from the physical world. By using AI to run thousands of virtual simulations or generate infinite design variations, firms can identify high-potential innovations faster and cheaper than traditional R&D methods allow [47]. This capacity to "fail fast" and iterate digitally leads to a higher volume of novel products and a faster time-to-market. Consequently, empirical studies indicate that the ability to use IS assets for exploration directly translates into superior innovation performance metrics, such as market share growth from new products and perceived novelty by customers [23,48].

**H5.** *AI-Enabled Exploration is positively associated with Innovation Performance.*

### *2.8. The Moderating Role of Competitive Pressure*

Finally, we introduce a boundary condition regarding the direct utilization of organizational resources. In our baseline model, we posit that Organizational Readiness impacts innovation primarily through the mediation of sensing capability. We do not hypothesize a direct link between readiness and exploration because resource-rich firms often suffer from organizational inertia [48]; simply having the budget and technology (readiness) does not automatically result in risky experimentation (exploration) if the market is stable.

This moderation does not contradict the mediation logic proposed earlier but rather introduces a theoretically distinct second mechanism. The serial mediation chain captures deliberate, sequential capability building: readiness enables sensing, which enables exploration. This is the default process under stable market conditions. However, contingency theory argues that high-velocity environments fundamentally alter how firms allocate resources [32]. When competitive pressure is acute, firms face time compression that makes the full sequential path untenable. Unable to wait for the deliberate sensing process to generate inputs for exploration, firms instead directly activate their slack resources for experimentation as a survival response [49, 50]. Gans, who examined artificial intelligence adoption in competitive markets, found that AI enhances long-term competitiveness by reducing costs and raising consumer surplus [51], further illustrating that competitive intensity compels firms to actively deploy their AI resources rather than allow them to remain latent. Competitive pressure therefore does not modify the mediation chain but rather activates a separate, parallel mechanism that bypasses it. These two mechanisms coexist in the model as complements: the mediated path explains capability-driven exploration under ordinary conditions, and the moderated direct path explains urgency-driven exploration under competitive threat. Conversely, in low-pressure environments, this direct relationship remains dormant.

**H6.** *Competitive Pressure positively moderates the relationship between Organizational Readiness and AI-Enabled Exploration, such that the relationship becomes significant and positive when pressure is high.*

Figure 1 illustrates the proposed conceptual framework. The model conceptualizes AI adoption not as a static event but as a dynamic process of capability development. On the left, technological (Technology Compatibility), organizational (Organizational Readiness), and environmental (Government Support) factors serve as the foundational antecedents. These factors do not directly generate innovation; rather, they facilitate the development of Sensing Capability, the firm's ability to scan and interpret market data. This sensing mechanism is posited as a necessary precursor to AI-Enabled Exploration, representing the active deployment of AI for experimentation and prototyping. Finally, this exploratory behavior is hypothesized to drive Innovation Performance. Additionally, the model incorporates Competitive Pressure as a critical boundary condition, moderating the translation of organizational resources into active exploration behavior.

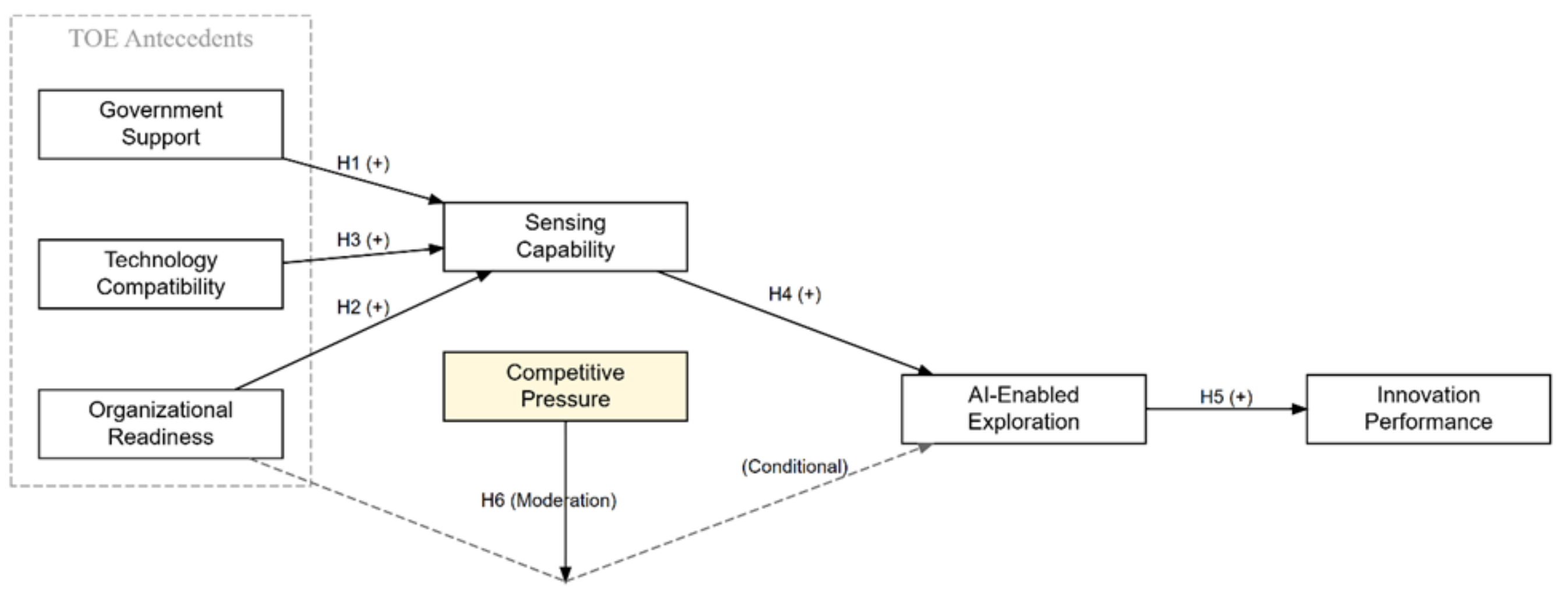

**Figure 1.** Proposed Conceptual Framework. Source: Authors' own elaboration.

## 3. Research Methodology

To empirically test the proposed research model, a field survey methodology was employed. This study employs a survey-based quantitative approach, which is appropriate as the study aims to capture the perceptions of organizational decision-makers regarding internal capabilities and environmental factors that are not directly observable through archival data [52]. The following subsections detail the research context, data collection procedures, measurement development, and analytical techniques.

### 3.1. Research Context

Innovation success rates in emerging markets are considerably lower than in advanced economies, making the study of contextual enablers and barriers particularly critical [53]. The empirical context for this study is the Kingdom of Saudi Arabia (KSA). This setting is particularly relevant for three reasons. First, under the Saudi Vision 2030 strategic framework, the nation is undergoing a rapid, state-led digital transformation, shifting from an oil-based economy to a knowledge-based economy driven by artificial intelligence and deep tech (KSA Vision 2030, 2016). Second, unlike Western economies where AI adoption is often market-driven, the Gulf Cooperation Council (GCC) region presents a unique environment where high Government Support and resource abundance coexist with organizational inertia, making it an ideal setting to study how firms overcome barriers to exploration. Third, despite the massive investment in AI infrastructure in the MENA region, empirical IS research remains heavily skewed toward North American and European contexts, leaving a gap in our understanding of AI-enabled dynamic capabilities in emerging high-growth economies.

These characteristics of the GCC institutional environment generate a specific empirical prediction. Under Vision 2030, government support for AI adoption has reached a point where it is available to nearly all firms simultaneously, creating low variance in the institutional stimulus across organizations. In such conditions, state support functions as a baseline condition rather than a differentiating resource, and internal factors such as readiness and compatibility are expected to carry greater explanatory weight than environmental support. This pattern distinguishes leapfrog economies from earlier-stage adopters where government support remains selectively distributed and therefore acts as a discriminating driver of capability development.

Interventionary studies involving animals or humans, and other studies that require ethical approval, must list the authority that provided approval and the corresponding ethical approval code.

In this section, where applicable, authors are required to disclose details of how generative artificial intelligence (GenAI) has been used in this paper (e.g., to generate text, data, or graphics, or to assist in study design, data collection, analysis, or interpretation). The use of GenAI for superficial text editing (e.g., grammar, spelling, punctuation, and formatting) does not need to be declared.

### 3.2. Data Collection and Sample

Data were collected from senior and middle managers in Saudi Arabian firms who possessed knowledge of their organization's IT strategy and innovation processes. To ensure a representative sample, we targeted a diverse range of industries including Telecommunications, Energy, Banking, and the Public Sector.

Given the cultural context of the region, where personal networks (Wasta) play a crucial role in business communication, we adopted a hybrid sampling strategy combining purposive sampling via professional networks (LinkedIn) and snowball sampling. Purposive sampling was employed to ensure that respondents possessed the organizational knowledge required to evaluate the constructs under study, specifically AI strategy, sensing capability, and innovation processes. Snowball sampling was adopted as a complementary strategy because, in high-context cultures such as Saudi Arabia, professional trust and personal referrals represent the primary gatekeeping mechanism for organizational access, making probability sampling impractical [54]. This approach is consistent with established practice in IS research conducted in GCC contexts, where network-based sampling has been shown to yield respondents with greater organizational knowledge and higher response quality than cold outreach methods. LinkedIn searches targeted senior and middle managers using job title filters including Chief Information Officer, IT Director, Innovation Manager, and Strategy Manager across the targeted industries. Initial invitations included a description of the study's purpose and eligibility criteria. Respondents who agreed to participate were subsequently asked to refer colleagues in similar managerial positions, with referrals limited to individuals meeting the same eligibility criteria to preserve sample quality. To ensure geographic and industry diversity, purposive quotas were applied across Riyadh, Jeddah, and the Eastern Province, and across the five targeted sectors.

A bilingual online survey (English and Arabic) was distributed. To ensure semantic equivalence, the instrument was translated into Arabic by a bilingual IS scholar and then back-translated into English by an independent linguist, following the procedure recommended by Brislin [55].

A total of 600 invitations were distributed. After a six-week data collection period, 268 responses were received. Following a rigorous data screening process, 23 responses were discarded due to incompleteness or straight-lining behavior (e.g., selecting "5" for all answers). This resulted in a final sample of 245 valid responses, yielding an effective response rate of 40.8%.

To assess potential nonresponse bias, we conducted a wave analysis following the procedure recommended by Armstrong and Overton [56]. This approach treats late respondents as proxies for non-respondents, on the assumption that those who respond later are more similar to those who did not respond at all. We divided the sample into early respondents (first 25%, $n = 61$) and late respondents (last 25%, $n = 61$) based on response timestamps. Independent samples t-tests comparing the two groups on all seven construct means revealed no statistically significant differences (all $p > 0.05$). Additionally, no significant differences were found between early and late respondents on key demographic variables including firm size, industry sector, and respondent job level. These results suggest that nonresponse bias is unlikely to pose a serious threat to the validity of the findings.

Table A1 (in Appendix A) presents the demographic profile of the respondents. The sample was well-distributed geographically, with the majority located in Riyadh (45%), Jeddah (30%), and the Eastern Province (25%). Firm size varied, with 35% from large enterprises (>500 employees) and 65% from SMEs, ensuring that the findings are generalizable across different organizational scales.

### 3.3. Measurements

To ensure content validity, measurement items were adapted from established scales in existing information systems and management literature. All items were measured on a 7-point Likert scale ranging from 1 ("Strongly Disagree") to 7 ("Strongly Agree"). Table 1 summarizes the constructs, their operational definitions in the context of this study, and the primary sources from which they were adapted.

**Table 1.** Definition of constructs and literature sources.

| Construct | Definition/Measurement Focus | Source |
|---|---|---|
| Government Support | The extent to which the government provides incentives, regulatory frameworks, and strategic initiatives to encourage technology adoption. | [57] |
| Organizational Readiness | The level of financial resources and technical infrastructure (preparedness) available to the firm for adopting advanced AI systems. | [11,37] |
| Technology Compatibility | The degree to which AI systems are perceived as consistent with the firm's existing values, needs, and past experiences. | [58] |
| Sensing Capability | The firm's ability to utilize AI tools to scan the external environment, detect market shifts, and identify new business opportunities. | [15] |
| AI-Enabled Exploration | The extent to which the firm uses AI systems to experiment with new business approaches and invent products/services that go beyond current offerings. | Adapted from [23] |
| Innovation Performance | The success of the firm in introducing new products and services, including the speed of response to market changes relative to competitors. | [59] |
| Competitive Pressure | The degree of intensity of competition within the industry that compels the firm to adopt AI to maintain strategic advantage. | [17] |

To establish content validity for the AI-Enabled Exploration items, we adapted each item from Jansen et al.'s [23] exploratory innovation scale, explicitly modifying the language to reference a specific AI mechanism: AI-driven insights for opportunity identification, Generative AI for invention, AI tools for business model experimentation, and AI simulations for commercialization. Following Hinkin [60], items adapted from established scales retain content validity when the adaptation is theoretically justified and subject to expert review. Consistent with the expert panel approach recommended by Lawshe [61] and the validation guidelines for IS research provided by Straub et al. [52], a pilot study was conducted with 15 IT executives and 5 academics to validate the face validity of the full instrument. Based on their feedback, minor wording adjustments were made to the Arabic version to ensure the items accurately reflected the local business context. The final list of measurement items is provided in Table A2 (in Appendix A).

### 3.4. Common Method Bias (CMB)

Since data for both independent and dependent variables were collected from a single source, Common Method Bias (CMB) could potentially inflate the relationships. We employed both procedural and statistical remedies to mitigate this risk [62]. Procedurally, respondents were assured of anonymity, and the order of questions was randomized to reduce consistency motifs.

Statistically, we performed three tests. First, we conducted Harman's single-factor test [63]. The exploratory factor analysis revealed that the first factor explained only 28.4% of the variance, well below the 50% threshold. Second, we applied the full collinearity assessment approach recommended by Kock [64], which treats inflated VIF values as indicators of common method variance rather than mere collinearity. All VIF values were below 3.0, providing evidence that CMB does not threaten the findings. Third, we employed the Common Latent Factor approach by introducing an unmeasured latent factor connected to all indicators in the CFA model. Adding the CLF did not improve model fit, indicating that common method variance does not systematically inflate the relationships in the model [65]. Furthermore, loading differences for the key theoretical constructs, AI-Enabled Exploration and Innovation Performance, were all below the 0.20 threshold (range: 0.015 to 0.040), confirming that the central path relationships are not biased by common method variance.

### 3.5. Data Analysis Technique

We utilized Partial Least Squares Structural Equation Modeling (PLS-SEM) to test the research model. PLS-SEM was chosen over covariance-based SEM (CB-SEM) for several reasons: (1) the research objective is prediction and theory development regarding the novel construct of AI-Enabled Exploration rather than confirmation of established theory [65]; (2) the model is complex, involving a serial mediation chain and a moderation interaction; and (3) PLS-SEM makes fewer assumptions regarding the normal distribution of data [66]. The analysis was performed using the seminr package in R, which allows for advanced bootstrapping and rigorous model estimation.

## 4. Results

To determine the nomological validity of the research model, a two-stage analytical approach was employed using Partial Least Squares Structural Equation Modeling (PLS-SEM). We utilized the seminr package in R (an advanced PLS-SEM library) to assess the measurement model by examining reliability and validity, and subsequently evaluated the structural model to test the hypothesized relationships. PLS-SEM was selected due to its suitability for exploratory research involving complex models with interaction terms and mediation effects [65].

### 4.1. Measurement Model Validation

Following standard procedures for PLS-SEM [67,68], we assessed the measurement model in two steps: indicator reliability and internal consistency, followed by convergent and discriminant validity. Given that AI-Enabled Exploration is a novel construct, we conducted additional validation tests including EFA and CFA to further corroborate its psychometric properties.

First, we examined the factor loadings of all indicator variables. As shown in Table 2, all retained items exhibited loadings well above the recommended threshold of 0.708, ranging from 0.827 to 0.953. One item from the Competitive Pressure construct (CP1) was removed due to low loading to ensure construct stability. Critically, all four AI-Enabled Exploration items returned particularly strong loadings ranging from 0.889 to 0.934, all significant at $p < 0.001$, confirming that each item reliably captures the intended construct.

**Table 2.** Loadings of the indicator variables (N = 245).

| Construct | Items | Loading | Significance |
|---|---|---|---|
| Government Support (GS) | GS1 | 0.892 | $p < 0.001$ |
| | GS2 | 0.924 | $p < 0.001$ |
| | GS3 | 0.827 | $p < 0.001$ |
| Organizational Readiness (OR) | OR1 | 0.932 | $p < 0.001$ |
| | OR2 | 0.923 | $p < 0.001$ |
| | OR3 | 0.913 | $p < 0.001$ |
| Technology Compatibility (TC) | TC1 | 0.853 | $p < 0.001$ |
| | TC2 | 0.879 | $p < 0.001$ |
| | TC3 | 0.918 | $p < 0.001$ |
| Sensing Capability (SC) | SC1 | 0.886 | $p < 0.001$ |
| | SC2 | 0.897 | $p < 0.001$ |
| | SC3 | 0.897 | $p < 0.001$ |
| AI-Enabled Exploration (AIE) | AIE1 | 0.920 | $p < 0.001$ |
| | AIE2 | 0.922 | $p < 0.001$ |
| | AIE3 | 0.934 | $p < 0.001$ |
| | AIE4 | 0.889 | $p < 0.001$ |
| Competitive Pressure (CP) | CP2 | 0.886 | $p < 0.001$ |
| | CP3 | 0.953 | $p < 0.001$ |
| Innovation Performance (IP) | IP1 | 0.847 | $p < 0.001$ |
| | IP2 | 0.881 | $p < 0.001$ |
| | IP3 | 0.834 | $p < 0.001$ |
| | IP4 | 0.869 | $p < 0.001$ |

Note: Item CP1 was dropped during scale purification.

Second, we evaluated construct reliability and convergent validity. As presented in Table 3, Cronbach's Alpha and Composite Reliability values for all constructs exceeded the 0.70 threshold, indicating high internal consistency. AVE for all constructs exceeded 0.50, confirming convergent validity. AI-Enabled Exploration returned particularly strong values, with Cronbach's Alpha of 0.936, Composite Reliability of 0.955, and AVE of 0.840, indicating that its items share substantially more variance with their intended construct than with measurement error.

Discriminant validity was assessed using two criteria. Using the Fornell-Larcker criterion [60], the square root of AVE for each construct exceeded its highest correlation with all other constructs. We further examined the HTMT ratio [70], considered a more rigorous test of discriminant validity. All HTMT values were well below the conservative threshold of 0.85. Notably, the highest observed value of 0.672 occurred between AI-Enabled Exploration and Sensing Capability, the two theoretically closest constructs in the model, yet this remains well below the threshold. Cross-loadings analysis further confirmed that all four AI-Enabled Exploration items loaded substantially higher on their intended construct (0.889 to 0.934) than on any other construct in the model, with the next highest loadings appearing on Sensing Capability (0.506 to 0.593) and Innovation Performance (0.534 to 0.571).

**Table 3.** Construct Reliability, Validity, and Discriminant Validity (Fornell-Larcker and HTMT).

| Construct | Cronbach's Alpha | Composite Reliability | AVE | GS | OR | TC | SC | AIE | CP | IP |
|---|---|---|---|---|---|---|---|---|---|---|
| Gov. Support (GS) | 0.864 | 0.913 | 0.778 | **0.882** | | | | | | |
| Org. Readiness (OR) | 0.913 | 0.945 | 0.851 | 0.044 | **0.923** | | | | | |
| Tech. Compat (TC) | 0.863 | 0.914 | 0.781 | 0.093 | 0.036 | **0.884** | | | | |
| Sensing Cap (SC) | 0.874 | 0.922 | 0.798 | 0.134 | 0.421 | 0.298 | **0.893** | | | |
| AI Exploration (AIE) | 0.936 | 0.955 | 0.840 | 0.066 | 0.409 | 0.198 | 0.672 | **0.916** | | |
| Comp. Pressure (CP) | 0.827 | 0.917 | 0.847 | 0.092 | 0.157 | 0.057 | 0.051 | 0.112 | **0.920** | |
| Innovation Perf (IP) | 0.881 | 0.918 | 0.736 | 0.112 | 0.311 | 0.052 | 0.410 | 0.664 | 0.219 | **0.858** |

Note: Bold diagonals represent the square root of average variance extracted (AVE). Values below the diagonal represent HTMT ratios.

To further establish the validity of the AI-Enabled Exploration scale, we conducted EFA and CFA analyses. A standalone EFA on the four AIE items confirmed a clean single-factor solution, with all items loading strongly on one factor (0.839 to 0.922), supported by excellent fit indices (RMSEA = 0.000, TLI = 1.007, $p$ = 0.977). Parallel analysis further corroborated this unidimensional structure. A 7-factor EFA on all items confirmed that AIE items loaded cleanly on their own factor (0.822 to 0.921) with negligible cross-loadings on all remaining factors. Finally, a full CFA of the measurement model returned strong fit indices (CFI = 0.970, TLI = 0.985, RMSEA = 0.003, SRMR = 0.032), with all AIE standardized loadings ranging from 0.840 to 0.918, all significant at $p$ < 0.001. Collectively, these results provide robust evidence that AI-Enabled Exploration is a psychometrically sound and empirically distinct construct.

*4.2. Structural Model and Hypothesis Testing*

Prior to assessing the path coefficients, we examined the Variance Inflation Factor (VIF) to check for collinearity issues. As reported in Table 4, all VIF values were below 3.0 (highest VIF = 1.197), indicating that multicollinearity is not a concern in this study [65].

We assessed the structural model by examining the coefficient of determination ($R^2$) and the significance of the path coefficients derived from a bootstrapping procedure with 5000 subsamples.

The model explained a substantial amount of variance in the endogenous constructs: 22.4% for Sensing Capability, 42.1% for AI-Enabled Exploration, and 36.5% for Innovation Performance. To assess out-of-sample predictive relevance, we conducted a PLSpredict analysis using 10 folds and 10 repetitions [71]. The PLS model consistently outperformed the linear model benchmark across all indicators for Sensing Capability, AI-Enabled Exploration, and Innovation Performance, returning lower RMSE and MAE values in all cases. $Q^2$ predict values were positive for all endogenous constructs (Sensing Capability = 0.029, AI-Enabled Exploration = 0.054, Innovation Performance = 0.013). This confirms that the model demonstrates strong predictive accuracy beyond the estimation sample. providing additional evidence of predictive relevance across all latent variables in the model.

The path analysis results are summarized in Table 4 and Figure 2. Regarding the antecedents (TOE factors), Organizational Readiness ($\beta = 0.376, t = 7.01, p < 0.001$) and Technology Compatibility ($\beta = 0.261, t = 5.17, p < 0.001$) were found to be significant positive predictors of Sensing Capability. However, contrary to expectations, Government Support ($\beta = 0.098, t = 1.65$) was not statistically significant ($p > 0.05$).

**Table 4.** Structural Model Results (Direct Effects).

| Hypothesis | Path Relationship | Path Coeff ($\beta$) | T-Statistic | Result | VIF | $f^2$ |
|---|---|---|---|---|---|---|
| H1 | Government Support →Sensing Cap | 0.098 | 1.65 | Not Supported | 1.009 | 0.012 |
| H2 | Org. Readiness→Sensing Cap | 0.376 *** | 7.01 | Supported | 1.000 | 0.182 |
| H3 | Tech. Compatibility→Sensing Cap | 0.261 *** | 5.17 | Supported | 1.009 | 0.087 |
| H4 | Sensing Cap→AI-Enabled Exploration | 0.530 *** | 12.30 | Supported | 1.174 | 0.412 |
| H5 | AI-Enabled Exploration→Innovation Perf | 0.604 *** | 15.44 | Supported | 1.000 | 0.576 |
| Control | Org. Readiness→AI-Enabled Exploration | 0.151 *** | 2.78 | (Main Effect) | 1.197 | - |
| | Comp. Pressure→AI-Enabled Exploration | 0.069 | 1.35 | (Main Effect) | 1.021 | - |
| Moderation | $OR \times CP$ →AI Exploration | 0.154 | 2.87 *** | Supported | 1.028 | 0.036 |

Note: *** $p < 0.001$.

Furthermore, Sensing Capability had a strong, positive, and significant effect on AI-Enabled Exploration ($\beta = 0.530, t = 12.30, p < 0.001$). Finally, AI-Enabled Exploration significantly influenced Innovation Performance ($\beta = 0.604, t = 15.44, p < 0.001$).

In addition to significance, we examined Cohen's $f^2$ effect sizes. As shown in Table 4, AI-Enabled Exploration had a large effect on Innovation Performance ($f^2 = 0.576$). Sensing Capability had a large effect on AI-Enabled Exploration ($f^2 = 0.412$), while Organizational Readiness had a medium effect on Sensing Capability ($f^2 = 0.182$). These effect sizes carry substantive practical implications. The large effect of AI-Enabled Exploration on Innovation Performance ($f^2 = 0.576$) indicates that a one standard deviation improvement in a firm's AI-enabled exploration capability is associated with a 0.604 standard deviation improvement in innovation outcomes, representing a meaningful return on investment in AI exploration infrastructure. Similarly, the large effect of Sensing Capability on AI-Enabled Exploration ($f^2 = 0.412$) indicates that building AI-driven sensing mechanisms is among the most impactful investments a firm can make on its path toward AI-driven innovation.

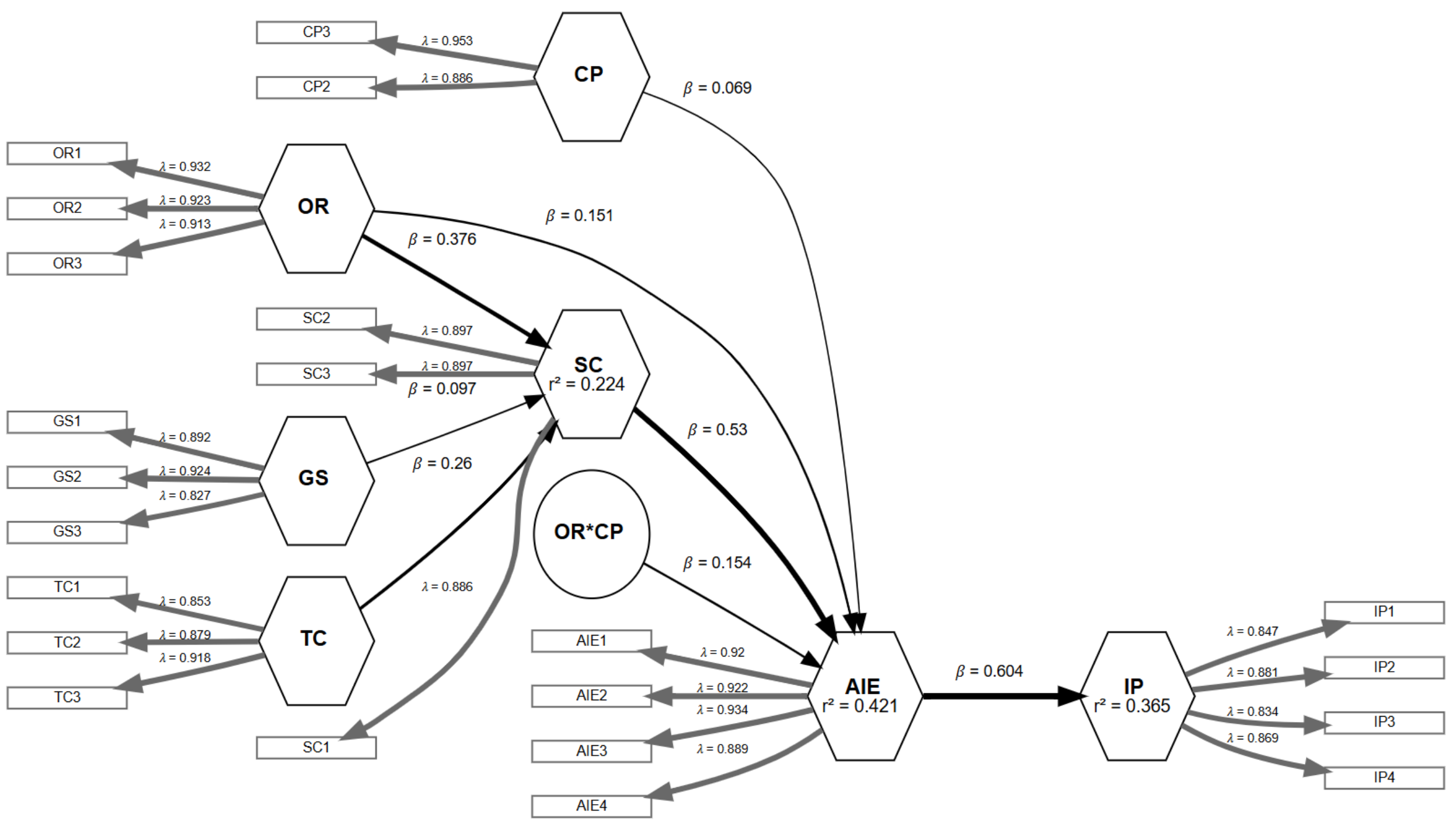

**Figure 2.** Structural Model Results (Path coefficients and $R^2$ values).

### 4.3. Mediation Analysis

To further understand the mechanisms through which capabilities translate into performance, we conducted a comprehensive analysis of indirect and total effects. As shown in Table 5, the serial mediation chain Organizational Readiness → Sensing Capability → AI-Enabled Exploration → Innovation Performance was positive and significant ($\beta = 0.120$, $t = 5.40$, $p < 0.001$), with a 95% confidence interval [0.080, 0.166] that did not include zero, confirming that the impact of organizational readiness on innovation is fully realized through the sequential development of sensing and exploration capabilities.

Beyond this primary mediation chain, we examined the indirect effect of Sensing Capability on Innovation Performance through AI-Enabled Exploration. This path was positive and significant ($\beta = 0.395$, $t = 9.907$, CI = [0.317, 0.474]), confirming that sensing capability's impact on innovation operates entirely through AI-Enabled Exploration rather than directly. This finding reinforces the sequential logic of the model: sensing capability identifies market opportunities, which AI-Enabled Exploration subsequently converts into innovation outcomes.

**Table 5.** Specific Indirect Effects (Mediation Analysis).

| Path Chain | Estimate | T-Stat | 95% CI [Lower, Upper] | Significance |
|---|---|---|---|---|
| SC → AIE → IP | 0.395 | 9.907 | [0.317, 0.474] | Yes |
| OR → SC → AIE → IP | 0.120 | 5.40 | [0.080, 0.166] | Yes |

Note: The arrow (→) denotes the directional path between constructs in the indirect effect chain (e.g., SC → AIE → IP indicates the effect of Sensing Capability on Innovation Performance transmitted through AI-Enabled Exploration).

To provide a complete picture of each construct's influence on Innovation Performance, Table 6 reports total effects decomposed into direct and indirect components. AI-Enabled Exploration exerts the strongest total effect ($\beta$ = 0.664), followed by Sensing Capability ($\beta$ = 0.395), Organizational Readiness ($\beta$ = 0.252), Technology Compatibility ($\beta$ = 0.115), and Government Support ($\beta$ = 0.041). These total effects confirm that the sequential capability-building chain substantially amplifies the influence of organizational resources on innovation beyond what is visible from direct paths alone.

**Table 6.** Total Effects on Innovation Performance.

| Path | Direct | Indirect | Total |
|---|---|---|---|
| GS → IP | 0.000 | 0.041 | 0.041 |
| TC → IP | 0.000 | 0.115 | 0.115 |
| OR → IP | 0.000 | 0.252 | 0.252 |
| SC → IP | 0.000 | 0.395 | 0.395 |
| AIE → IP | 0.664 | 0.000 | 0.664 |

"Note: The arrow (→) denotes the hypothesized directional effect of each construct on Innovation Performance (IP), decomposed into direct, indirect, and total effects." 4.4. Moderation Analysis

We hypothesized that Competitive Pressure would moderate the relationship between Organizational Readiness and AI-Enabled Exploration. As shown in Table 4, the interaction term ($OR \times CP$) was positive and significant ($\beta = 0.154, t = 2.87, p < 0.01$).

To interpret the nature of this interaction, we plotted the simple slopes analysis (see Figure 3). The plot reveals that the relationship between Readiness and Exploration is steeper for firms operating under High Competitive Pressure (Blue Line) compared to those under Low Competitive Pressure (Red Line). This indicates that while organizational readiness is a prerequisite for exploration, high market pressure acts as a catalyst, compelling firms to leverage their readiness more aggressively to explore new AI-driven opportunities.

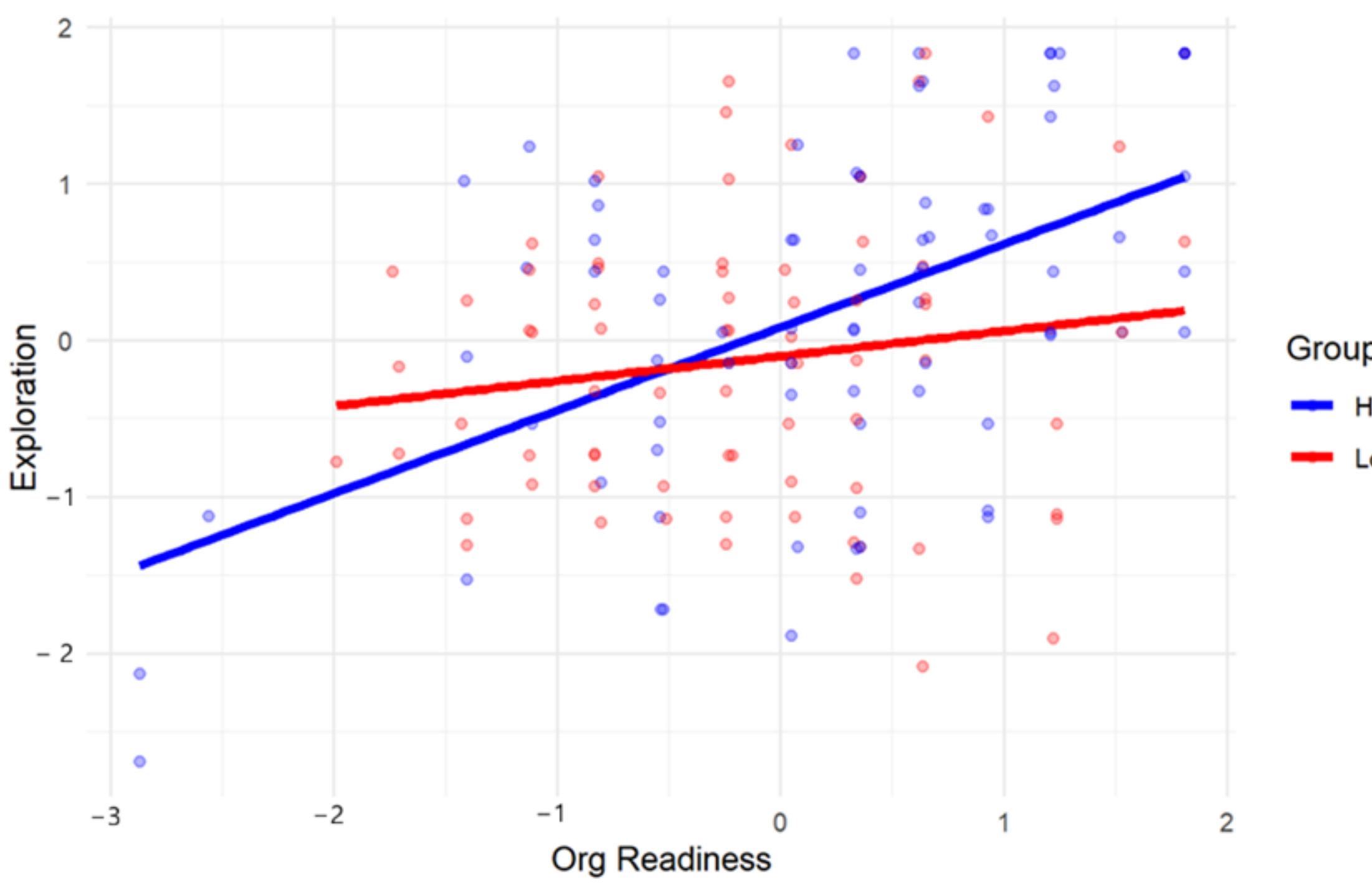

**Figure 3.** Moderating effect of Competitive Pressure on the relationship between Organizational Readiness and AI-Enabled Exploration. Blue dots represent the High Pressure group; red dots represent the Low Pressure group.

*4.5. Robustness and Sensitivity Analyses*

To further establish the methodological rigor of the findings, we conducted five additional analyses. First, we replicated the structural model using covariance-based SEM via the lavaan package in R. The CB-SEM results confirmed all directional and significance patterns from the PLS-SEM analysis: Government Support remained non-significant, Organizational Readiness and Technology Compatibility remained significant predictors of Sensing Capability, and AI-Enabled Exploration significantly predicted Innovation Performance. The CB-SEM model returned excellent fit indices (CFI = 0.990, TLI = 0.982, RMSEA = 0.003, SRMR = 0.041, Chi-sq/df = 0.974), confirming that the findings are not an artifact of the estimation method.

Second, we tested whether the core findings hold after introducing firm size and industry sector as control variables. All path coefficients remained substantively unchanged. The AI-Enabled Exploration to Innovation Performance coefficient was virtually identical with and without controls ($\beta = 0.604$ vs. $\beta = 0.605$), and $R^2$ improved marginally from 0.365 to 0.397, confirming that industry and firm size are not confounding the observed relationships.

Third, we addressed potential endogeneity concerns using the Gaussian copula approach [72]. The copula terms for all endogenous predictors were non-significant in both the AI-Enabled Exploration equation (SC_star: $p = 0.936$; OR_star: $p = 0.397$) and the Innovation Performance equation (AIE_star: $p = 0.713$), with $R^2$ remaining unchanged. These results indicate that endogeneity does not threaten the validity of our findings.

Fourth, a multigroup analysis was conducted to assess whether the model holds across firm size categories, comparing SMEs and large enterprises. No path coefficient showed a statistically significant difference between groups (all $p > 0.05$), confirming that the theoretical model generalizes across organizational scales.

Fifth, we compared our proposed serial mediation model against a rival direct effects model in which TOE factors predict Innovation Performance directly without mediation.

The proposed model explained 36.5% of variance in Innovation Performance compared to only 9.8% for the rival model, a difference of 26.7 percentage points. In the rival model, Government Support and Technology Compatibility were not significant predictors of performance. This comparison strongly validates the theoretical argument that dynamic capabilities serve as necessary mediating mechanisms between TOE antecedents and innovation outcomes.

## 5. Discussion

This study sets out to uncover the mechanisms through which firms in high-growth, state-guided economies leverage Artificial Intelligence to foster strategic exploration. Moving beyond the prevailing focus on AI for operational efficiency, we examined how technological, organizational, and environmental factors facilitate the development of sensing capability and AI-enabled exploration, and how these capabilities ultimately drive innovation performance. The empirical analysis of 245 Saudi Arabian firms largely supports our theoretical model, confirming that AI adoption is a multi-stage process of capability accumulation rather than a direct antecedent to performance.

Our findings reveal three critical insights into the nature of AI value creation, offering both confirmation of established theories and important deviations from prior empirical work. First, the results confirm the existence of a serial mediation chain wherein organizational readiness influences sensing capability, which in turn fosters AI-enabled exploration, ultimately leading to innovation performance. This validates our core theoretical argument that possessing AI resources alone is insufficient for innovation. Unlike prior studies that link IT assets directly to firm performance, our findings align with the process-oriented logic of the Dynamic Capabilities View, suggesting that financial and technical readiness must first be translated into a specific ability to sense market shifts before firms can effectively use AI to experiment with new solutions. This finding demystifies the "black box" of AI adoption, differentiating our work from recent studies such as Wong and Ngai, who focused primarily on operational performance [5]. We extend this line of inquiry by demonstrating that the path from investment to radical innovation is bridged by distinct, sequential capabilities focused on exploration rather than exploitation.

Second, regarding the antecedents of sensing capability, organizational readiness and technology compatibility emerged as strong, significant predictors. Government support, however, was not significant. This result is consistent with the institutional argument advanced in our research context. Because Vision 2030 initiatives have made state support effectively ubiquitous across Saudi firms, the construct captures a baseline condition rather than a differentiating resource. Empirical support for this interpretation comes from the near-zero correlation between government support and organizational readiness ($r = 0.044$, Table 3), indicating that state incentives are decoupled from the internal capability-building process. Support lowers the threshold for adoption but does not sort firms by capability level, which explains its non-significant effect on sensing capability. This pattern is consistent with institutional theory, which argues that when coercive pressures are uniformly applied across an organizational field, they drive isomorphic responses rather than differentiating behaviors [33]. Prior TOE research further supports this interpretation: Zhu and Kraemer demonstrated that environmental factors exert their strongest influence during initial adoption phases, gradually losing their explanatory power as technology becomes institutionally embedded across competing firms [17]. In the Saudi context, government support has effectively moved beyond this initial phase, functioning as an institutional background condition that enables adoption without shaping the depth of capability development. This finding adds nuance to the broader IS literature on institutional drivers by showing that their explanatory power is contingent on the stage of institutional maturation rather than being universally operative.

Third, our moderation analysis highlights the catalytic role of competitive pressure. We found that competitive pressure significantly strengthens the relationship between organizational readiness and AI-enabled exploration. This finding supports the "adapt-or-die" logic of contingency theory and refines previous assumptions that resource abundance automatically leads to innovation. In low-pressure environments, resource-rich firms may succumb to organizational inertia, hoarding resources without deploying them for exploration. However, our data indicates that when competitive intensity is high, firms are compelled to activate their latent readiness, leveraging AI to explore new frontiers aggressively to ensure survival. This nuances the Resource-Based View by demonstrating that external environmental conditions are necessary triggers to convert potential slack resources into active exploratory behavior.

### *5.1. Theoretical Contributions*

This study offers significant theoretical implications for the Information Systems and Management literature. Primarily, it extends the Dynamic Capabilities View by explicitly conceptualizing and providing initial empirical support for "AI-Enabled Exploration" as a distinct dynamic capability. While prior research by Pavlou and El Sawy extensively mapped the "sensing" and "seizing" capabilities, the specific role of Generative AI in facilitating the exploration of unknown domains has remained under-theorized [15]. By defining AI-enabled exploration as the organizational capacity to use generative algorithms for prototyping and experimentation, we distinguish it from traditional R&D processes. We provide empirical evidence that AI acts as a critical antecedent to organizational ambidexterity, enabling firms to overcome the resource constraints that typically force a difficult trade-off between exploitation and exploration.

Furthermore, we contribute to the TOE literature by refining the framework's application to emerging tech hubs. By demonstrating that environmental factors do not always act as direct drivers of capability, we challenge the assumption that institutional pressures universally translate into organizational competence. Our results suggest that in "leapfrog" economies like Saudi Arabia, external support acts merely as a baseline condition, while internal factors determine the depth of capability development. This adds complexity to the application of the TOE framework in state-guided economies, suggesting that future research must distinguish between environmental factors that trigger adoption versus those that sustain capability development.

Finally, by situating our research in Saudi Arabia, we provide rare empirical evidence of how "resource-rich, capability-poor" tensions are resolved in non-Western contexts. Most AI adoption studies originate from market-driven economies in North America or Europe. Our study shows that while financial capital is abundant in the region, the translation of that capital into innovation requires the distinct mechanism of AI-enabled sensing. This offers a theoretical roadmap for understanding digital transformation in other resource-rich nations attempting to transition to knowledge economies, filling a significant geographical gap in the global IS discourse.

Beyond its contributions to the IS and management literature, this study also offers implications for the emerging concept of Artificial Leadership [9]. By empirically demonstrating that AI-Enabled Exploration is preceded by a sensing mechanism, we show that AI-driven exploration is not a spontaneous organizational outcome but a capability that requires deliberate leadership investment in sensing infrastructure. This finding supports the Artificial Leadership argument that the transition from AI working 'for' leaders to AI working 'to' leaders is not a fixed threshold but a capability-dependent process. Organizations must first build the sensing capacity that allows AI to generate meaningful strategic inputs before AI can effectively serve as a source of exploratory ideas. The sequential sensing-to-exploration mechanism documented here thus provides an empirical foundation for understanding the organizational conditions that govern when and how firms can move from human-directed to AI-directed exploration.

### 5.2. Practical Implications

For practitioners, particularly managers in the MENA region, this study offers actionable guidance on investment priorities. Managers often fall into the trap of purchasing advanced AI tools without building the accompanying organizational processes. Our results show that readiness only leads to innovation if it passes through the mechanisms of sensing and exploration. Therefore, leaders should prioritize the establishment of dedicated AI units focused on scanning external data and running rapid digital prototypes before committing to large-scale operational rollouts. Furthermore, leaders in resource-rich firms should consider artificially inducing competitive pressure to overcome internal inertia, as our data shows that pressure is the essential trigger that turns potential resources into active exploration.

For policymakers, the finding that government support does not directly drive sensing capability suggests that the current policy focus on financial incentives may be reaching a point of diminishing returns. Policies that merely fund AI hardware may no longer be sufficient to drive distinct competitive advantages. Instead, policymakers should shift focus from funding adoption to fostering competitive markets and human capital development. Creating an environment of healthy market competition may be more effective at driving the actual usage of AI for innovation than direct subsidies, as competition forces firms to utilize their digital assets more effectively.

### 5.3. Limitations and Future Research

Despite its contributions, this study has limitations that present opportunities for future research. First, the data is cross-sectional, which captures a snapshot of AI adoption but precludes definitive causal claims about the long-term evolution of capabilities. Future research should employ longitudinal designs to track how AI-enabled exploration evolves over time as firms mature in their digital transformation. Second, while we focused on the positive outcome of innovation performance, we did not measure the potential downsides of AI exploration, such as the failure rate of AI-generated ideas or the costs associated with over-exploration. Future studies could investigate the dark side of AI experimentation to provide a more balanced view of its strategic impact.

## 6. Conclusions

This study addresses the pressing strategic imperative for organizations to transition from merely adopting Artificial Intelligence for operational efficiencies to leveraging it for genuine innovation and market exploration. By integrating the Technology–Organization–Environment framework with the Dynamic Capabilities View, we provide a comprehensive explanatory model that details how firms in high-growth, state-guided economies can cultivate the necessary internal mechanisms to harness the generative potential of AI. Our research challenges the deterministic view that AI investment automatically yields performance gains. Instead, we demonstrate that value creation is mediated by a specific sequence of dynamic capabilities, wherein organizational readiness enables the sensing of environmental shifts, which subsequently empowers the firm to engage in AI-enabled exploration.

The findings from the Saudi Arabian context offer a distinct contribution to the global discourse on digital transformation. We highlight that in resource-rich environments, financial and technical readiness are necessary but insufficient conditions for strategic renewal. The non-significant impact of government support suggests that while state initiatives effectively lower adoption barriers, they do not guarantee the development of sophisticated capabilities. Rather, it is the interplay of internal capability building and external competitive pressure that catalyzes the shift from inertia to active exploration. By providing preliminary empirical support for AI-enabled exploration as a critical mediator, this study charts a theoretical and practical pathway for firms attempting to navigate the uncertainties of the digital age. Taken together, these findings advance three theoretical boundaries simultaneously. First, they extend the Dynamic Capabilities View by identifying AI-Enabled Exploration as a micro-foundational seizing mechanism that bridges AI adoption and strategic renewal, offering a more granular theoretical vocabulary for understanding how AI systems generate value beyond automation. Second, they refine institutional theory's account of environmental drivers by demonstrating that coercive institutional pressures lose their differentiating power once they reach saturation, shifting the explanatory burden to internal organizational factors. Third, they enrich the ambidexterity literature by showing that AI specifically lowers the cost of the exploration side of ambidexterity, potentially allowing firms to pursue exploration without the structural redesign that prior research has shown to be necessary. Collectively, these contributions position AI not merely as a productivity tool but as a capability-building infrastructure whose strategic value is contingent on organizational conditions and competitive context, a perspective that has broad applicability beyond the Saudi setting to any economy where technology investment outpaces organizational readiness. It suggests that the future of competitive advantage lies not in the passive accumulation of digital assets, but in the active, AI-driven orchestration of knowledge to discover new frontiers of value.

**Author Contributions:** Conceptualization: S.N. and T.A., Research Design: S.N., Data Collection: T.A., Data Curation: T.A., Formal Data Analysis: S.N. and T.A., Writing—original draft: S.N. and T.A., Writing—review and editing: S.N. All authors have read and agreed to the published version of the manuscript.

**Funding:** This work was supported by the Deanship of Scientific Research, Vice Presidency for Graduate Studies and Scientific Research, King Faisal University, Saudi Arabia [Grant No KFU263308].

**Institutional Review Board Statement:** This study was approved by the Research Ethics Committee at King Faisal University (Approval Number: KFU-REC-2026-JAN-ETHICS3931).

**Data Availability Statement:** The dataset (CSV file) and the R script used for statistical analysis are available for replication from the corresponding author upon reasonable request.

**Conflicts of Interest:** The authors declare no conflicts of interest.

## Appendix A

**Table A1.** Profiles of the Respondents (N = 245). *Source: Authors' own elaboration.*

| Demographic Variable | Category | Frequency (*N*) | Percentage (%) |
|---|---|---|---|
| **Gender** | Male | 188 | 76.7% |
| | Female | 57 | 23.3% |
| **Age Group** | 20–29 years | 15 | 6.1% |
| | 30–39 years | 92 | 37.6% |
| | 40–49 years | 98 | 40.0% |
| | 50 years or older | 40 | 16.3% |
| **Education Level** | Bachelor's Degree | 135 | 55.1% |
| | Master's Degree | 96 | 39.2% |
| | Doctoral Degree | 14 | 5.7% |
| **Job Position** | C-Level Executive (CIO, CTO, CEO) | 42 | 17.1% |
| | IT Manager/Director | 98 | 40.0% |
| | Strategy/Innovation Manager | 65 | 26.5% |
| | Senior Technical Specialist | 40 | 16.3% |
| **Industry Sector** | Telecommunications & IT | 69 | 28.2% |
| | Energy & Manufacturing | 54 | 22.0% |
| | Banking & Finance | 44 | 18.0% |
| | Public Sector/Government | 49 | 20.0% |
| | Retail & Services | 29 | 11.8% |
| **Firm Location** | Riyadh (Central) | 110 | 44.9% |
| | Jeddah (West) | 74 | 30.2% |
| | Dammam/Khobar (East) | 61 | 24.9% |
| **Firm Size** | Small (< 50 employees) | 37 | 15.1% |
| | Medium (50–499 employees) | 122 | 49.8% |
| | Large (> 500 employees) | 86 | 35.1% |

**Table A2.** Measurement Items. *Source: Authors' own elaboration based on adapted scales.*

| Construct | Code | Measurement Item | Source |
|---|---|---|---|
| **Government Support** | GS1 | The government provides clear incentives for adopting AI technologies. | Adapted from [57] |
| | GS2 | There is adequate regulatory support for implementing AI in our industry. | |
| | GS3 | Government initiatives (e.g., Vision 2030) strongly encourage our digital transformation. | |
| **Organizational Readiness** | OR1 | We have the financial resources necessary to invest in advanced AI systems. | Adapted from[38] |
| | OR2 | We possess the technical infrastructure required to support Generative AI. | |
| | OR3 | Our employees have the necessary specialized skills to use AI tools effectively. | |
| **Technology Compatibility** | TC1 | Using AI systems is compatible with our current business operations. | Adapted from [58] |
| | TC2 | AI technology fits well with our organizational culture and values. | |
| | TC3 | Integrating AI into our workflow does not require major changes to our legacy systems. | |
| **Sensing Capability** | SC1 | We use AI-driven tools to frequently scan the environment for new business opportunities. | Adapted from [15] |
| | SC2 | Our AI systems allow us to review the effect of market changes on customers in real-time. | |
| | SC3 | We rely on AI analytics to identify new product development needs instantly. | |
| **AI-Enabled Exploration** | AIE1 | Our AI systems enable us to accept demands that go beyond our existing products and services. | Adapted from [23] |
| | AIE2 | We use Generative AI to invent new products and services that are completely new to our firm. | |
| | AIE3 | AI tools allow us to experiment with new business approaches in local markets. | |
| | AIE4 | We use AI simulations to commercialize products that are completely new to us. | |
| **Innovation Performance** | IP1 | Our new product introduction rate is higher than our major competitors. | Adapted from [59] |
| | IP2 | Our new products/services are often perceived as novel by customers. | |
| | IP3 | We are faster than competitors in responding to market changes with new innovations. | |
| | IP4 | Our innovations have significantly increased our market share. | |
| **Competitive Pressure** | CP1 * | We feel pressure from competitors who have already adopted AI. (*Dropped due to low loading*) | Adapted from [17] |
| | CP2 | Our main competitors are using AI to gain a strategic advantage. | |
| | CP3 | The adoption of AI is necessary to compete in our industry marketplace. | |

*Note: All items were measured on a 7-point Likert scale (1 = Strongly Disagree, 7 = Strongly Agree).*

* Dropped due to low loading

# References

1. Dong, X.; McIntyre, S.H. The Second Machine Age: Work, Progress, and Prosperity in a Time of Brilliant Technologies. *Quant. Financ.* **2014**, *14*, 1895–1896. https://doi.org/10.1080/14697688.2014.946440.
2. Davenport, T.H.; Ronanki, R. Artificial Intelligence for the Real World. *Harv. Bus. Rev. (HBR)* **2018**, *96*, 108–116.
3. Ali, M.; Khan, T.I.; Khattak, M.N.; Şener, İ. Synergizing AI and Business: Maximizing Innovation, Creativity, Decision Precision, and Operational Efficiency in High-Tech Enterprises. *J. Open Innov. Technol. Mark. Complex.* **2024**, *10*, 100352. https://doi.org/10.1016/J.JOITMC.2024.100352.
4. Alshurideh, M.T.; El Khatib, M.; Al Kurdi, B.; Nawaiseh, A.K.; Hamadneh, S.; Al-Sulaiti, K.; Alrawabdeh, W.A.; Alshurideh, A.; Shwedeh, F.; Alzoubi, H.M. Exploring the Impact of AI-Based Technology on Supply Chain Efficiency, with Mediator Role of Smart Inventory Management Practices. *Adv. Sci. Technol. Innov.* **2025**, *29*, 55–63. https://doi.org/10.1007/978-3-031-90131-7_7.
5. Wong, D.T.W.; Ngai, E.W.T. Impact of Artificial Intelligence (AI) on Operational Performance: The Role of Dynamic Capabilities. *Inf. Manag.* **2025**, *62*, 104162. https://doi.org/10.1016/J.IM.2025.104162.
6. Wang, J.; Shi, Y.; Jiang, X.; Venkatesh, V.G. How Does Artificial Intelligence Capacity Enhance the Production System Resilience and Operational Performance? A Human-Organization-Technology Fit Perspective. *Int. J. Inf. Manag.* **2026**, *87*, 103023. https://doi.org/10.1016/J.IJINFOMGT.2025.103023.
7. Raisch, S.; Birkinshaw, J. Organizational Ambidexterity: Antecedents, Outcomes, and Moderators. *J. Manag.* **2008**, *34*, 375–409. https://doi.org/10.1177/0149206308316058.
8. March, J.G. Exploration and Exploitation in Organizational Learning. *Organ. Sci.* **1991**, *2*, 71–87. https://doi.org/10.3280/SO2008-002006.
9. Kollmann, T.; Kollmann, K.; Kollmann, N. Artificial Leadership: Digital Transformation as a Leadership Task between the Chief Digital Officer and Artificial Intelligence. *Int. J. Bus. Sci. Appl. Manag.* **2023**, *18*, 76–95. https://doi.org/10.69864/ijbsam.18-1.172.
10. Dubey, R.; Gunasekaran, A.; Childe, S.J.; Blome, C.; Papadopoulos, T. Big Data and Predictive Analytics and Manufacturing Performance: Integrating Institutional Theory, Resource-Based View and Big Data Culture. *Br. J. Manag.* **2019**, *30*, 341–361. https://doi.org/10.1111/1467-8551.12355.
11. Mikalef, P.; Gupta, M. Artificial Intelligence Capability: Conceptualization, Measurement Calibration, and Empirical Study on Its Impact on Organizational Creativity and Firm Performance. *Inf. Manag.* **2021**, *58*, 103434. https://doi.org/10.1016/J.IM.2021.103434.
12. Wamba, S.F.; Gunasekaran, A.; Akter, S.; Ren, S.J.f.; Dubey, R.; Childe, S.J. Big Data Analytics and Firm Performance: Effects of Dynamic Capabilities. *J. Bus. Res.* **2017**, *70*, 356–365. https://doi.org/10.1016/J.JBUSRES.2016.08.009.
13. Benner, M.J.; Tushman, M.L. Exploitation, Exploration, and Process Management: The Productivity Dilemma Revisited. *Acad. Manag. Rev.* **2003**, *28*, 238–256. https://doi.org/10.5465/AMR.2003.9416096.
14. Gregory, R.W.; Keil, M.; Muntermann, J.; Mähring, M. Paradoxes and the Nature of Ambidexterity in IT Transformation Programs. *Inf. Syst. Res.* **2015**, *26*, 57–80. https://doi.org/10.1287/ISRE.2014.0554.
15. Pavlou, P.A.; El Sawy, O.A. Understanding the Elusive Black Box of Dynamic Capabilities. *Decis. Sci.* **2011**, *42*, 239–273. https://doi.org/10.1111/J.1540-5915.2010.00287.X.
16. Teece, D.J. Profiting from Innovation in the Digital Economy: Enabling Technologies, Standards, and Licensing Models in the Wireless World. *Res. Policy* **2018**, *47*, 1367–1387. https://doi.org/10.1016/J.RESPOL.2017.01.015.
17. Zhu, K.; Kraemer, K.L. Post-Adoption Variations in Usage and Value of e-Business by Organizations: Cross-Country Evidence from the Retail Industry. *Inf. Syst. Res.* **2005**, *16*, 61–84. https://doi.org/10.1287/ISRE.1050.0045.
18. Al-Somali, S.A.; Gholami, R.; Clegg, B. An Investigation into the Acceptance of Online Banking in Saudi Arabia. *Technovation* **2009**, *29*, 130–141. https://doi.org/10.1016/J.TECHNOVATION.2008.07.004.
19. Dwivedi, Y.K.; Kshetri, N.; Hughes, L.; Slade, E.L.; Jeyaraj, A.; Kar, A.K.; Baabdullah, A.M.; Koohang, A.; Raghavan, V.; Ahuja, M.; et al. Opinion Paper: "So What If ChatGPT Wrote It?" Multidisciplinary Perspectives on Opportunities, Challenges and Implications of Generative Conversational AI for Research, Practice and Policy. *Int. J. Inf. Manag.* **2023**, *71*, 102642. https://doi.org/10.1016/J.IJINFOMGT.2023.102642.
20. Warner, K.S.R.; Wäger, M. Building Dynamic Capabilities for Digital Transformation: An Ongoing Process of Strategic Renewal. *Long Range Plan.* **2019**, *52*, 326–349. https://doi.org/10.1016/J.LRP.2018.12.001.
21. Tornatzky, L.G.; Fleischer, M. *The Processes of Technological Innovation*; Lexington Books: Lexington, KY, USA, 1990.
22. Teece, D.J. Explicating Dynamic Capabilities: The Nature and Microfoundations of (Sustainable) Enterprise Performance. *Strateg. Manag. J.* **2007**, *28*, 1319–1350. https://doi.org/10.1002/SMJ.640.

23. Jansen, J.J.P.; Van Den Bosch, F.A.J.; Volberda, H.W. Exploratory Innovation, Exploitative Innovation, and Performance: Effects of Organizational Antecedents and Environmental Moderators. *Manag. Sci.* **2006**, *52*, 1661–1674. https://doi.org/10.1287/MNSC.1060.0576.
24. He, Z.L.; Wong, P.K. Exploration vs. Exploitation: An Empirical Test of the Ambidexterity Hypothesis. *Organ. Sci.* **2004**, *15*, 481–495. https://doi.org/10.1287/ORSC.1040.0078.
25. Teece, D.J.; Pisano, G.; Shuen, A. Dynamic Capabilities and Strategic Management. *Strateg. Manag. J.* **1997**, *18*, 509–533. https://doi.org/10.1002/(SICI)1097-0266(199708)18:7<509::AID-SMJ882>3.0.CO;2-Z.
26. Ferreira, J.; Coelho, A. Dynamic Capabilities, Innovation and Branding Capabilities and Their Impact on Competitive Advantage and SME's Performance in Portugal: The Moderating Effects of Entrepreneurial Orientation. *Int. J. Innov. Sci.* **2020**, *12*, 255–286. https://doi.org/10.1108/IJIS-10-2018-0108.
27. Kowalski, M.; Bernardes, R.C.; Gomes, L.; Borini, F.M. Microfoundations of Dynamic Capabilities for Digital Transformation. *Eur. J. Innov. Manag.* **2025**, *28*, 3717–3746. https://doi.org/10.1108/EJIM-12-2023-1074.
28. Hwang, B.N.; Lai, Y.P.; Wang, C. Open Innovation and Organizational Ambidexterity. *Eur. J. Innov. Manag.* **2023**, *26*, 862–884. https://doi.org/10.1108/EJIM-06-2021-0303.
29. Nambisan, S.; Lyytinen, K.; Majchrzak, A.; Song, M. Digital Innovation Management: Reinventing Innovation Management Research in a Digital World. *MIS Q.* **2017**, *41*, 223–238. https://doi.org/10.25300/MISQ/2017/41:1.03.
30. Nevi, G.; Pizzichini, L.; Bastone, A.; Dezi, L. Adoption of AI by Micro and Small Health Enterprises: Effects of Entrepreneurial Orientation on the TOE Model. *Eur. J. Innov. Manag.* **2025**, *28*, 209–241. https://doi.org/10.1108/EJIM-07-2024-0770.
31. Ben Slimane, S.; Coeurderoy, R.; Mhenni, H. Digital transformation of small and medium enterprises: A systematic literature review and an integrative framework. *Int. Stud. Manag. Organ.* **2022**, *52*, 96–120.
32. Zhu, K.; Kraemer, K.L.; Xu, S. The Process of Innovation Assimilation by Firms in Different Countries: A Technology Diffusion Perspective on e-Business. *Manag. Sci.* **2006**, *52*, 1557–1576. https://doi.org/10.1287/MNSC.1050.0487.
33. Lawrence, P.R.; Lorsch, J.W. Differentiation and Integration in Complex Organizations. *Adm. Sci. Q.* **1967**, *12*, 1–47. https://doi.org/10.2307/2391211.
34. Dimaggio, P.J.; Powell, W.W. The Iron Cage Revisited: Institutional Isomorphism and Collective Rationality in Organizational Fields. *Am. Sociol. Rev.* **1983**, *48*, 147–160. https://doi.org/10.2307/2095101.
35. Teo, H.H.; Wei, K.K.; Benbasat, I. Predicting Intention to Adopt Interorganizational Linkages: An Institutional Perspective. *MIS Q.* **2003**, *27*, 19–49. https://doi.org/10.2307/30036518.
36. Baker, J. The Technology–Organization–Environment Framework. In *Information Systems Theory: Explaining and Predicting Our Digital Society*; Springer: New York, NY, USA, 2012; pp. 231–245. https://doi.org/10.1007/978-1-4419-6108-2_12.
37. Barney, J. Firm Resources and Sustained Competitive Advantage. *J. Manag.* **1991**, *17*, 99–120.
38. Iacovou, C.L.; Benbasat, I.; Dexter, A.S. Electronic Data Interchange and Small Organizations: Adoption and Impact of Technology. *MIS Q.* **1995**, *19*, 465–485. https://doi.org/10.2307/249629.
39. Ben Slimane, S. Absorption–technological absorptive capacity and innovation: The primacy of knowledge. In *Innovation Economics, Engineering and Management Handbook 2: Special Themes*; John Wiley & Sons: Hoboken, NJ, USA, 2021; pp. 43–49.
40. Mikalef, P.; Krogstie, J.; Pappas, I.O.; Pavlou, P. Exploring the Relationship between Big Data Analytics Capability and Competitive Performance: The Mediating Roles of Dynamic and Operational Capabilities. *Inf. Manag.* **2020**, *57*, 103169. https://doi.org/10.1016/J.IM.2019.05.004.
41. Rogers, E.M.; Singhal, A.; Quinlan, M.M. *Diffusion of Innovations*; Free Press: New York, NY, USA, 1995; ISBN 9781351358712.
42. Dedrick, J.; West, J. Why firms adopt open source platforms: A grounded theory of innovation and standards adoption. In Proceedings of the Workshop on Standard Making: A Critical Research Frontier for Information Systems, Seattle, WA, USA, 12–14 December 2003.
43. Zabel, C.; O'Brien, D. Understanding Dynamic Capabilities in Emerging Technology Markets: Antecedents, Sequential Nature, and Impact on Innovation Performance in the Extended Reality Industry. *Eur. J. Innov. Manag.* **2024**, *27*, 305–336. https://doi.org/10.1108/EJIM-07-2023-0574.
44. Overby, E.; Bharadwaj, A.; Sambamurthy, V. Enterprise Agility and the Enabling Role of Information Technology. *Eur. J. Inf. Syst.* **2006**, *15*, 120–131. https://doi.org/10.1057/PALGRAVE.EJIS.3000600.
45. Roberts, N.; Grover, V. Leveraging Information Technology Infrastructure to Facilitate a Firm's Customer Agility and Competitive Activity: An Empirical Investigation. *J. Manag. Inf. Syst.* **2012**, *28*, 231–270. https://doi.org/10.2753/MIS0742-1222280409.
46. Lin, Y.K.; Maruping, L.M. Organizing for AI Innovation: Insights From an Empirical Exploration of U.S. Patents. *Manag. Inf. Syst. Q.* **2025**, *49*, 1095–1122. https://doi.org/10.25300/MISQ/2025/18765.

47. Thomke, S.H. *Experimentation Matters: Unlocking the Potential of New Technologies for Innovation*; Harvard Business School Press: Boston, MA, USA, 2003.
48. Joshi, K.D.; Chi, L.; Datta, A.; Han, S. Changing the Competitive Landscape: Continuous Innovation through IT-Enabled Knowledge Capabilities. *Inf. Syst. Res.* **2010**, *21*, 472–495. https://doi.org/10.1287/ISRE.1100.0298.
49. Hannan, M.T.; Freeman, J. Structural Inertia and Organizational Change. *Am. Sociol. Rev.* **1984**, *49*, 149–164. https://doi.org/10.2307/2095567.
50. Bourgeois, L.J. On the Measurement of Organizational Slack. *Acad. Manag. Rev.* **1981**, *6*, 29–39. https://doi.org/10.2307/257138.
51. Gans, J.S. Artificial Intelligence Adoption in a Competitive Market. *Economica* **2023**, *90*, 690–705. https://doi.org/10.1111/ECCA.12458.
52. Straub, D.; Boudreau, M.-C.; Gefen, D. Validation Guidelines for IS Positivist Research. *Commun. Assoc. Inf. Syst.* **2004**, *13*, 24. https://doi.org/10.17705/1CAIS.01324.
53. Singh, S. How the Digital Environment Moderates Disruptive Technology and Digital Entrepreneurship Relationship in Emerging Markets. *Eur. J. Innov. Manag.* **2025**, *28*, 2719–2732. https://doi.org/10.1108/EJIM-04-2024-0426.
54. Saunders, M.; Lewis, P.; Thornhill, A. *Research Methods for Business Students*, 8th ed.; Pearson: Harlow, UK, 2019.
55. Brislin, R.W. Back-Translation for Cross-Cultural Research. *J. Cross. Cult. Psychol.* **1970**, *1*, 185–216. https://doi.org/10.1177/135910457000100301.
56. Armstrong, J.S.; Overton, T.S. Estimating Nonresponse Bias in Mail Surveys. *J. Mark. Res.* **1977**, *14*, 396–402. https://doi.org/10.2307/3150783.
57. Tan, M.; Teo, T.S.H. Factors Influencing the Adoption of Internet Banking. *J. Assoc. Inf. Syst.* **2000**, *1*, 5. https://doi.org/10.17705/1jais.00005.
58. Moore, G.C.; Benbasat, I. Development of an Instrument to Measure the Perceptions of Adopting an Information Technology Innovation. *Inf. Syst. Res.* **1991**, *2*, 192–222. https://doi.org/10.1287/ISRE.2.3.192.
59. Lai, F.; Li, D.; Wang, Q.; Zhao, X. The Information Technology Capability of Third-Party Logistics Providers: A Resource-Based View and Empirical Evidence from China. *J. Supply Chain Manag.* **2008**, *44*, 22–38. https://doi.org/10.1111/J.1745-493X.2008.00064.X.
60. Hinkin, T.R. A Brief Tutorial on the Development of Measures for Use in Survey Questionnaires. *Organ. Res. Methods* **1998**, *1*, 104–121.
61. Lawshe, C.H. A Quantitative Approach to Content Validity. *Pers. Psychol.* **1975**, *28*, 563–575.
62. Podsakoff, P.M.; MacKenzie, S.B.; Lee, J.Y.; Podsakoff, N.P. Common Method Biases in Behavioral Research: A Critical Review of the Literature and Recommended Remedies. *J. Appl. Psychol.* **2003**, *88*, 879–903. https://doi.org/10.1037/0021-9010.88.5.879.
63. Harman, H.H. *Modern Factor Analysis*; University of Chicago Press: Chicago, IL, USA, 1976.
64. Kock, N. Common Method Bias in PLS-SEM: A Full Collinearity Assessment Approach. *Int. J. E-Collab.* **2015**, *11*, 1–10. https://doi.org/10.4018/IJEC.2015100101.
65. Hair, J.F.; Risher, J.J.; Sarstedt, M.; Ringle, C.M. When to Use and How to Report the Results of PLS-SEM. *Eur. Bus. Rev.* **2019**, *31*, 2–24. https://doi.org/10.1108/EBR-11-2018-0203.
66. Chin, W.W. The Partial Least Squares Approach to Structural Equation Modeling. In *Modern Methods for Business Research*; Marcoulides, G.A., Ed.; Lawrence Erlbaum Associates: Mahwah, NJ, USA, 1998; pp. 295–336.
67. Anderson, J.C.; Gerbing, D.W. Structural Equation Modeling in Practice: A Review and Recommended Two-Step Approach. *Psychol. Bull.* **1988**, *103*, 411–423. https://doi.org/10.1037/0033-2909.103.3.411.
68. Gefen, D.; Straub, D.; Boudreau, M.-C. Structural Equation Modeling and Regression: Guidelines for Research Practice. *Commun. Assoc. Inf. Syst.* **2000**, *4*, 7. https://doi.org/10.17705/1CAIS.00407.
69. Fornell, C.; Larcker, D.F. Evaluating Structural Equation Models with Unobservable Variables and Measurement Error. *J. Mark. Res.* **1981**, *18*, 39–50.
70. Henseler, J.; Ringle, C.M.; Sarstedt, M. A New Criterion for Assessing Discriminant Validity in Variance-Based Structural Equation Modeling. *J. Acad. Mark. Sci.* **2014**, *43*, 115–135. https://doi.org/10.1007/S11747-014-0403-8.
71. Shmueli, G.; Sarstedt, M.; Hair, J.F.; Cheah, J.H.; Ting, H.; Vaithilingam, S.; Ringle, C.M. Predictive Model Assessment in PLS-SEM: Guidelines for Using PLSpredict. *Eur. J. Mark.* **2019**, *53*, 2322–2347.
72. Park, S.; Gupta, S. Handling Endogenous Regressors by Joint Estimation Using Copulas. *Mark. Sci.* **2012**, *31*, 567–586.